\documentstyle[aps]{revtex}
\begin{document}
\newcommand{\beq}{\begin{equation}}
\newcommand{\eeq}{\end{equation}}
\newcommand{\beqn}{\begin{eqnarray}}
\newcommand{\eeqn}{\end{eqnarray}}
\newcommand{\bmath}{\begin{mathletters}}
\newcommand{\emath}{\end{mathletters}}

\twocolumn[\hsize\textwidth\columnwidth\hsize\csname @twocolumnfalse\endcsname 

\title{Slope of the superconducting gap function 
in $Bi_2Sr_2CaCu_2O_{8+\delta}$ measured by vacuum tunneling 
spectroscopy}
\author{J. E. Hirsch }
\address{Department of Physics, University of California, San Diego\\
La Jolla, CA 92093-0319}
 
\date{\today} 
\maketitle 
\begin{abstract} 
Reproducible scanning tunneling microscope (STM) spectra of 
$Bi_2Sr_2CaCu_2O_{8+\delta}$ consistently exhibit asymmetric tunneling
characteristics, with the higher peak conductance corresponding to a 
negatively biased sample. We consider various  possible sources of this
asymmetry that are not intrinsic to the superconducting state, 
including energy dependence of the normal state densities
of states of sample and/or tip, existence of bandwidth cutoffs, unequal work 
functions of tip and sample, and energy-dependent transmission probability.
It is concluded that none of these effects can explain the sign and
temperature dependence of the observed asymmetry. This indicates that the
observed asymmetry
reflects an intrinsic property of the superconducting state: an
energy-dependent superconducting gap function with non-zero slope at
the Fermi energy. It is pointed out that
such a sloped gap function will also give rise to 
a thermoelectric effect in STM experiments, resulting in a $positive$
thermopower. We discuss the feasibility
of observing this thermoelectric effect with an STM and conclude that
it is easily observable. Again, contributions to this thermoelectric effect
may also arise from energy dependence of normal state
densities of states and from energy-dependent transmission probability.
However, because each of these features manifests itself differently in the
thermoelectric effect and in the tunneling characteristics, an analysis of
thermoelectric currents and voltages together with the tunneling spectra as 
function of temperature and tip-sample distance would allow for 
accurate determination of the slope of the gap function. It is suggested
that it would be very worthwhile to perform these experiments, because the slope
of the gap function reflects a fundamental property of the
superconducting state. In particular, the theory of hole superconductivity 
has predicted the existence of such a slope, of universal sign, in all superconductors.
It is furthermore argued that recent experimental results on vortex lattice
imaging provide further strong evidence for the existence of the gap slope discussed
here.

\end{abstract}
\vskip2pc]

\section{Introduction}

Tunneling spectroscopy is a powerful experimental tool that has
the potential to provide fundamental information on the 
microscopic physics of high temperature superconductivity.
Unfortunately, due to the difficulty in preparation of well
characterized sample surfaces and tunnel junctions, often the
spectra obtained have shown varying features that depend on the
particular sample and/or the particular tunneling technique used.
Consequently it has been difficult to determine which of the 
features found in the large number of experimental studies that
have been performed to date actually reflect intrinsic
properties of high temperature superconductors.

It is hence notable that several recent scanning tunneling
microscopy (STM) studies of single crystals of $Bi_2Sr_2CaCu_2O_{8+\delta}$
cleaved in ultrahigh vacuum have shown highly reproducible
features.\cite{stm1,stm2,stm3} These experiments are performed 
under carefully controlled conditions,
and tunneling spectra are obtained at a large number of different
positions on the sample and for a range of distances between tip
and sample. The spectra do not show time dependence
and generally exhibit either superconducting or semiconducting
features, which has been suggested to depend on the oxygen
stochiometry of the topmost Cu-O layer at the position of the
tip\cite{stm1}.

The spectra which exhibit superconducting features show structure 
in the tunneling conductance
$dI/dV$ that resembles the structure in the tunneling density of
states of conventional BCS superconductors predicted by 
BCS theory: a decrease in the conductance for low voltages, where
a gap-like structure develops, and a large enhancement of the
conductance at the edges of this gap-like structure. A notable
difference with the predictions of conventional BCS theory is that the 
spectrum is highly asymmetric: the peaks in $dI/dV$ are symmetrically
located around zero voltage but their height is substantially different,
with the largest peak corresponding to a negatively biased sample.
This asymmetry is found to persist to low temperatures.
Another feature of the spectra, which is sometimes observed only for
negative sample voltage\cite{stm1}, 
and sometimes for both polarities\cite{stm3}, is a dip
in the conductance at voltages about twice the position of the 
conductance peaks.

In this paper we focus on the tunneling asymmetry
in the peaks of $dI/dV$ and
consider various possible origins for it. Firstly, it is clear that
the tunneling geometry is asymmetric itself; an asymmetry may be
expected to arise from different work functions of the tip and 
the sample. In fact, tunneling asymmetries are often seen in
conventional tunnel junctions and have been explained assuming
differences in work functions of the electrodes\cite{dyne}. Secondly,
as we will show, a tunneling asymmetry  also arises from the fact
that the tunneling probability is energy-dependent, even in the
absence of differences in work functions. Finally, a tunneling 
asymmetry could also arise from an energy-dependent normal state density
of states of either tip or sample, or from the existence of a
bandwidth cutoff in either tip or sample. To focus the discussion
we assume that the superconducting state in the sample is described
by
conventional BCS theory with an isotropic gap, so that the 
density of states in the superconducting state is
\beq
N(E(\epsilon))=g_s(\epsilon)\frac{E}{\sqrt{E^2-\Delta^2}}
\eeq
with $g_s
(\epsilon)$ the density of states of the sample
in the normal state, and the quasiparticle energy $E$ given in
terms of the band energy $\epsilon$ by
\beq
E(\epsilon)=\sqrt{(\epsilon-\mu)^2+\Delta^2}
\eeq
with $\Delta$ the energy gap and $\mu$ the chemical potential.
We will also allow for the possibility of broadening of the
density of states arising from lifetime or other effects\cite{dyne2},
characterized by an inverse lifetime $\Gamma$.

We will show that none of the
effects listed above is able to describe even qualitatively
an asymmetry of the sign and temperature dependence seen 
experimentally. Briefly, asymmetries originating in
asymmetric normal state densities of states or bandwidth 
cutoffs tend to become smaller as the temperature is lowered,
contrary to experimental observations. Asymmetry originating in
energy dependence of the transmission probability persists
down to low temperatures but is of opposite sign to that observed
experimentally. Asymmetry originating in differences in
work functions can give the correct sign of the observed asymmetry
at high temperature but the sign switches as the temperature is
lowered, in contradiction with experiment.

We are thus led to the conclusion that the observed asymmetry 
originates in an intrinsic feature of the superconducting state.
 Such asymmetry will arise when the
superconducting gap function $\Delta$ is itself energy dependent\cite{mars1},
so that the quasiparticle energy Eq. (2) is
\beq
E(\epsilon)=\sqrt{(\epsilon-\mu)^2+\Delta(\epsilon)^2} .
\eeq
The measured asymmetry reflects the slope of the gap function
at the Fermi energy, $\Delta '(\mu)$. Hence we argue that the
experimental results are direct evidence for the existence of a
slope in the superconducting gap function.

Another experimental consequence of a sloped gap function is that  
it should manifest itself in a thermoelectric effect in tunneling experiments.\cite{hirs1} 
It gives rise to a thermoelectric power of the tunnel junction given 
by $Q=\Delta(\mu) \Delta '(\mu)/eT$ for small temperature gradients. We 
argue that such a thermoelectric effect should be easily observable
with an STM and that it should be experimentally searched for. Certainly,
energy dependence of the normal state densities of states and energy-dependent
transmission could also contribute to a thermoelectric effect. 
However we will show that analysis of the thermoelectric currents and
voltages in conjunction with the tunneling characteristics $dI/dV$ as
function of temperature and tip-sample distance should allow to clearly distinguish
between the different sources of a thermoelectric effect,
 and in particular to accurately determine the slope of the gap function. 
The sign of the observed tunneling asymmetry implies that the sign of the
measured thermoelectric power will be $positive$ (at least for small tip-sample
distance), opposite to what one would predict from energy dependence of
the electron transmission probability alone. In fact, thermoelectric experiments
with STM with normal metal electrodes have
recently been reported\cite{ther1}, with the sign of the measured themopower
always negative.

The slope of the gap function at the
Fermi energy is a fundamental property of the superconducting state,
as it gives information on the underlying microscopic interactions
that give rise to superconductivity in the system. For example, for an
attractive Hubbard model\cite{micn} or for an electron-phonon model within a
local approximation (Holstein model)\cite{free}
 the superconducting gap function has zero slope.
The detection of a gap slope would reveal the existence of an intrinsic electron-hole
asymmetry of the superconductor and shed light on the theoretical understanding
of superconductivity. In particular, the theory of hole superconductivity\cite{hole1}
has predicted the
existence of such a slope, of universal sign, in all superconductors, with magnitude
that scales with the critical temperature. However, a sloped superconducting
gap function could also arise in other theoretical frameworks.

It is currently generally believed that high $T_c$ superconductors have an order
parameter with complete or dominant d-wave symmetry\cite{harl}. The d-wave symmetry
relates to the variation of the gap $on$ the Fermi surface, while the
slope of the gap function discussed here relates to variation of the gap in direction
$perpendicular$ to the Fermi surface. Here we assume a gap that is constant
on the Fermi surface for simplicity; however a simple generalization of
our analysis could also be
used to  describe the variation of the amplitude of a d-wave gap,
or more generally of an anisotropic gap,
in direction perpendicular to the Fermi surface, and we believe the conclusions
would be similar concerning the existence of an $average$ gap slope. 
Thus we argue that our analysis is relevant irrespective of
whether the gap has s-wave or d-wave or any other symmetry. 

We note also that recent 
work\cite{coff} has argued that a gap with d-wave symmetry together with an 
assumption of directional tunneling can explain the experimentally
observed tunneling asymmetries. We will discuss that work in the conclusions.

In the next section we discuss the tunneling formalism. Sect. III deals with
barrier effects, and Sect. IV with density of states effects. In Sect. V
we consider the effect of an energy-dependent gap function. Section VI discusses
the expected thermoelectric effect in the presence of all these effects. We conclude 
in Sect. VII with a discussion. 

\section{Formalism}

The tunneling probability for an electron of energy $E$ above the
Fermi level of the tip is given within the WKB approximation by
\bmath
\beq
T(E,V)=e^{-2S(E,V)}
\eeq
\beq
S(E,V)=\int_0^d\sqrt{\frac{2m}{\hbar^2}(V(x)-E)}
\eeq
\emath
with $m$ the electron mass and $V(x)$ the barrier potential.
We assume a trapezoidal barrier, with work functions $\phi_t$ and
$\phi_s$ for tip and sample respectively, and a voltage $V$ of the
sample relative to the tip, as shown in Fig. 1, and obtain
\beqn
S(E,V)&=&\frac{\sqrt{2m}}{\hbar}\frac{2d}{3(\phi_t+V-\phi_s)}\times\nonumber\\
& &[(\phi_t-E)^{3/2}-(\phi_s-V-E)^{3/2}]
\eeqn
for the case where both the arguments raised to the $3/2$ power are
positive. When the voltage or energy are such that one of the
arguments is negative, Eq. (5) still holds if that argument is
set equal to zero. If the work functions $\phi_t$ and $\phi_s$ are
similar and the voltage $V$ is small Eq. (5) reduces to
\beq
S(E,V)=\frac{\sqrt{2m}}{\hbar}d(\bar{\phi}-\frac{V}{2}-E)^{1/2}
\eeq
with $\bar{\phi}=(\phi_s+\phi_t)/2$ the average work function.
The derivation assumes $V$, $V+E<<\bar{\phi}$ and
$|\phi_t-\phi_s|<<\bar{\phi}$.

We assume that electrons tunneling into the superconductor go into
a band of width $D$. Within a BCS formalism the tunneling current from
sample to tip is given by\cite{tink}
\beqn
I_{st}(V)&=&\int_{-D/2}^{D/2}d\epsilon [u^2(\epsilon)
[f_t(E(\epsilon)-eV)-f_s(E(\epsilon))]\nonumber\\
&\times &g_t(E(\epsilon)-eV)T(E(\epsilon)-eV,V)\nonumber\\
&+&v^2(\epsilon)[f_s(E(\epsilon))-f_t(E(\epsilon)+eV)]\nonumber\\
&\times &g_t(-E(\epsilon)-eV)T(-E(\epsilon)-eV,V)]g_s(\epsilon)
\eeqn
Here, $f_t$ and $f_s$ denote Fermi functions for tip and sample, which
may be at different temperatures, and $g_t$ and $g_s$ are normal state
densities of states for tip and sample. 
$V$ is the voltage
of the sample relative to the tip. The coherence factors are given
by the usual form
\bmath
\beq
u^2(\epsilon)=\frac{1}{2}(1+\frac{\epsilon-\mu}{E(\epsilon)})
\eeq
\beq
v^2(\epsilon)=\frac{1}{2}(1-\frac{\epsilon-\mu}{E(\epsilon)})
\eeq
\emath
and we assume that the relation between quasiparticle energy $E$ and
band energy $\epsilon$ is given by Eq. (2). $\mu$ is the sample 
chemical potential.

The tunneling current Eq. (7) can be written as
\beqn
I_{st}(V)&=&\int_{-\infty}^{\infty}d\omega
[f_t(\omega-eV)-f_s(\omega)]\nonumber\\
&\times &g_t(\omega-eV)T(\omega-eV,V)\rho_s(\omega)
\eeqn
with $\rho_s(\omega)$ the superconducting density of states
\beq
\rho_s(\omega)=\int_{-D/2}^{D/2}d\epsilon [u^2(\epsilon)\delta(\omega-E(\epsilon))
+v^2(\epsilon)\delta(\omega+E(\epsilon))]g_s(\epsilon).
\eeq
Broadening effects\cite{dyne2} may be included by replacing
the $\delta$-functions in Eq. (10) by
\beq
\delta(\omega)\rightarrow\frac{1}{\pi}\frac{\Gamma}{\omega^2+\Gamma^2}
\eeq
with $\Gamma$ a phenomenological broadening parameter.

We consider for definiteness a band of width $D=0.5eV$ in the sample,
a critical temperature $T_c=100K$ and zero temperature gap
$\Delta(T=0)=30meV$. The temperature dependence of the gap is assumed
to follow the conventional weak coupling BCS behavior. Figure 2 shows
for reference the tunneling conductance obtained from Eq. (7) assuming unit
transmission and energy-independent normal state densities of states
of sample and tip, for broadening factors $\Gamma=0$ and
$\Gamma=3meV$. The amplitude of the peaks in $dI/dV$ is
proportional to $u^2$ and $v^2$ for positive and negative voltage
respectively. Because 
$u^2(\epsilon)=v^2(2\mu-\epsilon)$,
tunneling characteristics are symmetric in the sign of $V$ in the absence of 
other effects.

\section{Barrier effects}
When we allow for energy-dependent transmission as given by
Eq. (4), the tunneling spectra develop an asymmetry that increases
as the temperature is lowered, as shown in Fig. (3a). By decreasing the
work functions of tip and sample or increasing the barrier thickness 
the size of the asymmetry is increased. Thus
it would easily be possible to match the magnitude and 
temperature-dependence of the observed asymmetry. However, the sign of the asymmetry
thus obtained, larger conductance for positively biased sample, is
opposite to what is seen experimentally.

At first sight the existence of this asymmetry may seem puzzling, since
it occurs for equal work functions of tip and sample and for densities
of states that are symmetric around the Fermi energy for both tip and
sample.
However it is easily understood by inspection of the diagram in Fig. 4. Because  
the gap in the superconducting density of states of the sample 
blocks the transmission
in that energy range, the transmission for positive sample involves electrons 
that are closer to the top of the barrier than that for a negative sample.
As the temperature is raised and the gap in the density of states 
closes, this asymmetry will disappear.

Could energy-dependent transmission give rise to an asymmetry of the observed
sign? It could happen two ways. First, if the transmission coefficient for
an electron were to decrease as its energy increases. While this cannot happen in
an STM, it could happen in a conventional tunnel junction if the Fermi level
is in the bandgap of the insulating barrier close to the valence band, as
discussed by Gundlach \cite{gund}. Second, energy-dependent transmission
would give rise to an asymmetry of the observed sign if the gap structure
in the density of states shown in Fig. 4 occurred in the tip rather than in the sample.
Clearly, neither of these scenarios appears to be relevant to the case under
consideration.

Next we consider the effect of differences in the work function of tip and sample. 
This can give rise to asymmetry in tunneling, and in fact has been 
proposed to explain asymmetries observed in various conventional tunnel 
junctions\cite{dyne}.
If the work function of the tip is lower than that of the sample the tunneling
conductance in the normal state will be larger for a negatively biased sample,
in agreement with observations. Figure 3(b) shows results for a rather extreme case, with
work functions $\phi_t=0.1eV$, $\phi_s=3eV$. While the asymmetry in the normal
state is of the observed sign, as the temperature is lowered the peaks in 
the conductance exhibit the asymmetry of opposite sign, induced by the energy
dependence of the transmission in the presence of the superconducting 
energy gap. It is found
that no combination of work functions and barrier thicknesses can
give rise to an asymmetry of the observed sign in the peaks of the tunneling
conductance at low temperatures.

\section{Density of states effects} 

We next consider the effect of non-constant density of states and of bandwidth cutoffs.
Figure 5 shows the resulting conductance when the normal state density
of states of the sample and/or of the tip has a linear energy variation,
reflected in the conductance for $T=T_c$ in Figure 5. This is obtained for
an electronic density of states in the tip that increases with energy,
and/or one in the sample that decreases with energy. As the temperature is
lowered it is seen that the magnitude of the asymmetry in the conductance
peaks decreases, simply because the peaks arise from band energies
increasingly closer to the Fermi energy. When we include broadening, as in
Fig. 5(b), the asymmetry persists to lower temperatures, but there is
still a clear progressive reduction in the magnitude of the asymmetry as
the temperature decreases. 

Furthermore, when we include the energy dependence of the transmission the
asymmetry gets further reduced and may even change sign, as shown in Fig. 6.
In (a) the normal state asymmetry, of the sign seen experimentally,
 arises from a sloped density of states
of either the tip or the sample. In (b), we assume that the band edge
in the sample occurs very close to the Fermi energy, as shown by the
dashed line, also giving rise to an asymmetry of the sign seen experimentally
at high temperatures.\cite{nega} In the presence of energy-dependent transmission, 
the asymmetry changes sign as the temperature is lowered.
The work functions assumed here are not particularly small, and smaller work
functions or larger tip-sample distances would make the obtained asymmetry 
deviate even further from what is observed.

We conclude that asymmetries originating from non-constant densities of
states and bandwidth cutoffs cannot explain the experimentally observed
asymmetry. The asymmetry originating from these effects invariably decreases
rapidly as the temperature is lowered. The asymmetry originating in
energy dependence of transmission discussed in the previous section 
does persist to low temperatures but it is of opposite sign to the one
observed experimentally. Hence we conclude that the experimentally 
observed asymmetry must originate in an intrinsic asymmetry of the 
superconducting state itself.

\section{Energy-dependent gap function}

The simplest generalization of BCS theory with constant energy gap is to
assume that the gap depends on wavevector $k$ only through the band energy
$\epsilon_k$. Hence the quasiparticle energy is
\beq
E_k(\epsilon_k)=\sqrt{(\epsilon_k-\mu)^2+\Delta^2(\epsilon_k)}.
\eeq
For energies close to the Fermi energy,
\beq
\Delta(\epsilon)=\Delta(\mu)+\Delta'(\mu)(\epsilon-\mu).
\eeq
The minimum quasiparticle energy will now occur not at $\epsilon_0=\mu$
but rather at
\beq
\epsilon_0=\mu-\Delta(\mu)\Delta'(\mu)
\eeq
and is
\beq
\Delta_0=E(\epsilon_0)=\Delta(\mu)+O(\Delta'^2).
\eeq
The BCS coherence factors Eq. (8) are no longer equal at the band
energy corresponding to the minimum quasiparticle energy, $\epsilon_0$; rather,
\bmath
\beq
u^2(\epsilon_0)=\frac{1}{2}(1-\Delta')
\eeq
\beq
v^2(\epsilon_0)=\frac{1}{2}(1+\Delta')
\eeq
\emath
with $\Delta'\equiv\Delta'(\mu)$.
For electrons injected into (extracted from) the sample,
i.e. positive (negative) sample voltage,  the tunneling
conductance is proportional to $u^2$ ($v^2$). Hence if the slope $\Delta'$ is positive,
that is if the gap function is an increasing function of electronic energy,
the tunneling current and conductance will be larger for a negatively biased
sample, as seen experimentally. We define
\beq
\frac{dI}{dV})_+=height\ of\ \frac{dI}{dV}\ peak\ for\ V_{sample}>0
\eeq
and similarly for $V_{sample}<0$, and the tunneling asymmetry as
the difference in the height of the peaks divided by the
average peak height
\beq
A=\frac{\frac{dI}{dV})_- - \frac{dI}{dV})_+}
{[\frac{dI}{dV})_ + + \frac{dI}{dV})_+]/2}
\eeq
and we have at low temperatures\cite{mars1} 
\beq
A(T\rightarrow0)=2\Delta'(\mu)
\eeq
Figure 7 shows the tunneling conductance for gap slope $\Delta'=0.12$, with
and without broadening. The asymmetry is clearly noticeable in both
cases and approaches the theoretical limit Eq. (19) for the case of
no broadening, while it remains somewhat smaller for $\Gamma \neq 0$.

However the effect of the gap slope needs to be balanced with
that of the energy-dependent transmission. Expanding the expression
Eq. (5) for the transmission we obtain
\bmath
\beq
T(E,V)=T_0[1+c(E-eV/2)]
\eeq
\beq
c(eV^{-1})=0.51\frac{d(A)}{\bar{\phi}(eV)^{1/2}}
\eeq
\emath
with $\bar{\phi}$ the average of the work functions of tip and 
sample. We find for the tunneling asymmetry at low temperatures
\beq
A(T\rightarrow0)=2(\Delta'(\mu)-\frac{c\Delta_0}{2}).
\eeq
For a typical barrier $\bar{\phi}=1eV$, $d=10A$ we have $c\sim5eV^{-1}$.
Hence for a gap $\Delta_0=30meV$ the tunneling asymmetry will be of
the observed sign if the gap slope is larger than about $0.08$.

The slope of the gap function, $\Delta '$, may or may not depend
on temperature. If $\Delta '$ is independent of $T$, then its effect will become
increasingly important relative to the effect of energy-dependent transmission
as the temperature is raised, according to Eq. (21), since
$\Delta_0$ decreases as $T_c$ is approached. For example, for
parameters where the intrinsic and barrier-induced asymmetry exactly cancel
at low temperatures, the intrinsic asymmetry would dominate as $T$
approaches $T_c$. Instead, if the gap slope scales with the gap itself,
its effect on the asymmetry will have the same temperature dependence as
that of energy-dependent transmission. We will assume
that this is the case here, so that the gap slope 
approaches zero as $T\rightarrow T_c$. It should also be noted that this is
the behavior predicted within
the theory of hole superconductivity\cite{hole1}. 

Figure 8 shows the combined effects of intrinsic asymmetry and
energy dependent transmission, for a case where intrinsic asymmetry
dominates: $\Delta'=0.12$, and barrier parameters
$\phi_s=\phi_t=1eV$, $d=10A$, leading to $A=0.044$ in Eq. (21).
In the presence of also a non-constant density of states, the asymmetry seen in
the normal state persists strongly as the temperature is lowered,
Fig. 8(b), unlike the cases where no intrinsic asymmetry exists (Figs. 5 and 6).

Eqs. (20) and (21) imply that the observed asymmetry should become
weaker as the tip-sample distance increases and the effect of 
energy-dependent transmission increases. There is some evidence for
this effect in the data shown in Fig. 13 of Ref. 2. The asymmetry for
the smallest tip-sample distance (largest conductance) is approximately
$12.5 \%$, and it decreases to about $5-7\%$ as the tip-sample 
distance increases and the current decreases by about a factor of $10$.
The change in the tip-sample distance over that range will be 
approximately
\beq
\Delta d(A)=\frac{ln10}{1.02\sqrt{\bar{\phi}(eV)}}
\eeq
From Eqs. (20)-(22) we may deduce the values of the average work
function and the change in the tip-sample distance in the experiment
of Ref. 2. Assuming that the asymmetry decreased from $12.5\%$ to $6\%$
when the current decreased by a factor of $10$ and a gap $\Delta_0=30meV$ yields
\bmath
\beq
\bar{\phi}=0.53 eV
\eeq
\beq
\Delta d=3.1 A
\eeq
\emath
which are not unreasonable. The gap slope implied by these data is
\beq
\Delta'=0.062+0.011d(A)
\eeq
where $d$ is the smallest tip-sample distance in the
data in Fig. 13 of Ref. 2. For example, $\Delta'=0.17$ if $d=10A$. This
analysis furthermore implies that the observed asymmetry will switch sign if the
tip-sample distance were to be increased by another $3A$, corresponding to
another factor of $10$ drop in the conductance. In Figure 9 we show
the tunneling characteristics for these parameters at low
temperatures for a range of tip-sample distances illustrating this
effect. It would be of interest
to obtain more detailed experimental information on the 
dependence of the tunneling asymmetry on tip-sample separation to compare
with the theoretical expectation.

We conclude from this analysis that it is possible to account
for the observed tunneling asymmetry in a simple way by assuming
an energy-dependent gap function with a finite slope at the Fermi energy.
The existence of energy-dependent transmission does not qualitatively
modify the results, it only implies that the underlying gap slope giving
rise to the observed asymmetry is 
larger than what one would have inferred assuming the transmission was energy-independent.
To explain the energy dependence of the observed spectra above $T_c$
it is furthermore necessary to assume that there is either energy dependence in
the normal state density of states and/or a large difference in the 
work functions of tip and sample. However, these factors would not be able
to explain the persistence of the asymmetry to low temperatures.
Accurate measurement of the 
dependence of the tunneling asymmetry on tip-sample distance at
low temperatures should
allow for a determination of the effect of energy-dependent transmission
and hence for accurate determination of the gap slope. As we show in the
next section, further information on the gap slope may be inferred by
measurement of a thermoelectric effect.

\section{Thermoelectric effect in STM}

Further information on the existence of a slope in the gap function may
be obtained from experiments where the tip and the sample are
at different temperatures. To our knowledge such experiments at cryogenic
temperatures
have not been yet been systematically performed\cite{japa}, but we are not
aware of any fundamental reason that would not allow for thermally decoupling of 
tip and sample and keeping them at substantially different temperatures in vacuum. 
Assuming there is no heat conduction path between tip and sample,
the only way to transport heat is through radiation. For example,
assume the sample is at higher temperature than the tip. The power 
transfered from sample to tip per unit area is of order
\bmath
\beq
P=\sigma (T_s^4 - T_t^4)
\eeq
\beq
\sigma=5.67 10^{-5}erg sec^{-1}cm^{-2}K^{-4}
\eeq
\emath
To maintain a temperature difference  between tip and sample the tip needs to
conduct away the heat it absorbs sufficiently fast. A typical metallic 
conductivity is $\kappa \sim 10^{7}erg sec^{-1}cm^{-1} K^{-1}$, so the
temperature gradient developed in the tip will be 
\beq
\nabla T\sim \frac{\sigma}{\kappa}(T_s^4-T_t^4)\sim \frac{10^{-11}}
{cmK^3}(T_s^4-T_t^4)\eeq
For example, for a tip of $1cm$ length at $T=0$ at the end far from the
sample, the end close to the sample would heat up to a negligible $T_t\sim 10^{-3}K$
if the sample is at temperature $T_s=100K$. A detailed analyisis of 
heat transfer between tip and sample in an STM is given in Ref. 19.

There have been in fact several recent investigations of thermoelectric effects
in scanning tunneling microscopy with normal metal 
samples\cite{ther1,ther2,ther3,ther4,ther5}. In these experiments, performed
at room temperature, the tip is heated by laser power to a temperature
approximately $10K$ higher than the sample. It is noteworthy that in these
experiments the thermopower measured is invariably negative, even when the
samples are metals with positive bulk values for the thermopower.
This is presumably due to the contribution to the thermopower from 
energy-dependent transmission, as discussed below.

If we assume that the gap slope $\Delta '$ is positive, as indicated by 
the tunneling asymmetry, a positive current will flow from the hotter
to the colder electrode in our case 
in the absence of other effects, that is, the system will
have positive thermopower. This is a consequence of the fact that the average
quasiparticle charge in the superconductor, given by 
\beq
q(\epsilon)=v^2(\epsilon)-u^2(\epsilon)
\eeq
is positive on the average\cite{hirs1}, 
unlike the usual BCS case with constant energy gap where 
quasiparticles are charge neutral on average.

For energy-independent transmission and constant densities of states
the current from sample to tip originating from quasiparticles
of lowest energy, i.e. with $E=\Delta_0$, is proportional to
\beqn
J_{st}&=&(1-\Delta')[f_t(\Delta_0-eV)-f_s(\Delta_0)]\nonumber\\
&+&(1+\Delta')[f_s(\Delta_0)-f_t(\Delta_0+eV)]\nonumber\\
&=&f_t(\Delta_0-eV)-f_t(\Delta_0+eV)\nonumber\\
&-&\Delta'[
f_t(\Delta_0-eV)+f_t(\Delta_0+eV)-2f_s(\Delta_0)]
\eeqn
and for small temperature gradient this yields for the zero current
thermoelectric voltage of the sample
\beq
V_0=
\frac{\Delta_0}{e}\Delta'\frac{T_s-T_t}{T_t}
\eeq
hence the thermopower of the junction, $Q$, is directly 
proportional to the gap slope:
\beq
Q=\frac{\Delta_0}{eT_t}\Delta'
\eeq
and is positive if the gap slope is positive. Note that the thermopower here 
increases as the temperature is lowered, unlike the
usual situation in metals and semiconductors where it decreases
linearly with temperature as $T\rightarrow 0$.

However, contributions to the thermopower will also arise from
energy-dependent transmission as well as from non-constant densities
of states. If we consider the current arising only from quasiparticles
of minimum energy, $\Delta_0$, a non-constant density of states of the
sample will not contribute at low temperatures. Eq. (28) generalizes to
\beqn
J_{st}&=&(1-\Delta')(1+c\Delta_0)(1+c_t\Delta_0)
[f_t(\Delta_0-eV)-f_s(\Delta_0)]\nonumber\\
&+&(1+\Delta')(1-c\Delta_0)(1-c_t\Delta_0)[f_s(\Delta_0)-f_t(\Delta_0+eV)]\nonumber\\
& &
\eeqn
assuming that the voltage $V<<\Delta_0$. Here, $c_t$ is the logarithmic
derivative of the tip density of states at the Fermi energy
\beq
c_t=g'_t(0)/g_t(0)
\eeq
and we have approximately
\beqn
J_{st}&=&f_t(\Delta_0-eV)-f_t(\Delta_0+eV)
-[\Delta'-\Delta_0(c+c_t)]\nonumber\\
&\times &[f_t(\Delta_0-eV)+f_t(\Delta_0+eV)-2f_s(\Delta_0)]
\eeqn
which contributes to the zero voltage thermoelectric current proportionally to
\beq
J_{st}=\frac{\Delta_0}{e}(\Delta'-\Delta_0(c+c_t))[f_t(\Delta_0)-f_s(\Delta_0)]
\eeq
For the zero current thermoelectric voltage we have for small
temperature gradient
\beq
V_0=
\frac{\Delta_0}{e}[\Delta'-\Delta_0(c+c_t)]\frac{T_s-T_t}{T_t}
\eeq
Hence we conclude that the thermoelectric voltage may be of either
sign depending on the magnitude of the parameters
$c$ and $c_t$ arising from energy-dependent transmission and
tip density of states respectively.

In particular, note that the effect of energy-dependent transmission
is twice as large here as it is in the tunneling asymmetry, Eq. (21). This is because the
tunneling asymmetry occurs at finite voltage, $V=\Delta_0$, where the effect
of energy-dependent transmission is smaller than at zero voltage. Hence it is
possible to have a tunneling asymmetry of the observed sign, reflecting the
intrinsic gap slope, together with a negative thermopower, dominated by
energy-dependent transmission.

It is also possible to obtain a negative thermopower just from the effect
of tip density of states if $c_t$ is positive, that could dominate over the effect of the
intrinsic gap slope. However, note that this corresponds to a tip density of states
that $increases$ with electronic energy, and hence it would give rise to
tunneling spectra with normal state slope that is opposite to the slope
that is observed experimentally.

Figures 10 and 11 illustrate some of the effects discussed above. Figure 10
shows current versus voltage in the presence of a large temperature
difference between sample and tip, with the tip colder than the sample. The
thermoelectric effect arising from intrinsic asymmetry and from a 
sloped tip density of states $g_t$ is similar, as seen in Fig. 10
(curves labeled $a$ and $b$), and it corresponds
to positive thermoelectric power. However the $dI/dV$ spectra (Figs. 11a and 11b) are
very different, with the ones corresponding to the sloped $g_t$ exhibiting an asymmetry
that is opposite in sign to that observed experimentally and to that originating
in the intrinsic effect, Fig. 11a. Conversely, a sloped $g_t$ that would give
rise to an asymmetry of the sign seen experimentally would give rise to a thermopower
opposite in  sign to what is shown in Fig. 10, i.e. negative thermopower. Thus
experimental results for tunneling spectra together with thermoelectric
effect could clearly distinguish between competing hypothesis of
intrinsic asymmetry and energy-dependent $g_t$.

An energy-dependent $g_s$ of negative slope would give rise to an asymmetry in 
tunneling of the sign seen experimentally (Fig. 11c) and to a thermoelectric
effect that has the same sign as that given by the intrinsic asymmetry, but the
magnitude of the thermoelectric current is much smaller than in the other
cases, as seen in Fig. 10 (curve labeled $c$). The zero voltage thermoelectric
current arising from all these effects is given by
\beqn
I_{ts}&=&\frac{2}{eR}\int_{\Delta_0}^\infty
dE\frac{E}{\sqrt{E^2-\Delta_0^2}}\times\nonumber\\
&[&\frac{\Delta'\Delta_0}{E}
+(c+c_t)E+c_s\frac{E^2-\Delta_0^2}{E}][f_t(E)-f_s(E)]
\eeqn
wit $c_s=g'_s(0)/g_s(0)$.
The effect of non-constant sample density of states is much smaller
because the singularity at $E=\Delta_0$ is cancelled for the term 
involving $c_s$. Thus it would be possible to discern the effect
of intrinsic asymmetry versus sample density of states on the thermopower
by calculating the expected magnitude of thermoelectric
current and comparing with experiment.

The energy-dependent transmission gives rise to a tunneling asymmetry
of opposite sign to that seen experimentally, as previously discussed,
and to negative thermopower. However, because its effect on the 
thermopower is twice as large as on the tunneling asymmetry it
is possible to choose parameters so that the tunneling asymmetry is 
of the sign observed experimentally, as determined by $\Delta'$, and yet 
the thermopower is negative, opposite to what the intrinsic asymmetry
predicts. An example is shown in Figs. 10 (curve labeled $d$)
 and 11d. However this requires
rather large values of both intrinsic asymmetry and energy-dependent
transmission parameters.

Analysis of the thermoelectric effect as function of temperature
would also allow to clearly distinguish between an intrinsic
origin and the other effects. Figure 12 shows the zero voltage
thermoelectric current versus sample temperature $T_s$ for a fixed
temperature difference between sample and tip, $T_t-T_s=-0.1$, with
the tip colder than the sample. In the presence of only the
intrinsic asymmetry (solid line) the thermoelectric current would go to zero at 
$T_c$. Instead, energy-dependent density of states of either tip
or sample (dashed and dotted lines) can give a thermoelectric current of the same
sign, but it continues to increase as the temperature is raised above
$T_c$. The same is true for the current arising from energy-dependent
transmission, which in addition is of opposite sign to that originating
in intrinsic asymmetry (dash-dotted line). Results for these cases for temperature gradient
of opposite sign, $T_t-T_s=0.1$, are the mirror image across the
horizontal axis of these results.

In contrast, the magnitude of the zero current thermoelectric voltage
$V_0$ is rather different for both signs of the temperature gradient,
 being larger when the tip is colder than the sample.
At low temperatures we can derive an expression for $V_0$ that is
not restricted to small temperature gradients:
\beq
V_0=[\Delta'-\Delta_0(c+c_t)]\frac{k_BT_t}{e}
[e^{(\beta_t-\beta_s)\Delta_0}-1]
\eeq
which shows that the thermoelectric voltage can become very
large at low temperatures when the tip is colder than the sample. 
Again, the voltage goes to zero at $T_c$ if it originates in
intrinsic asymmetry and remains finite when it originates
in the other factors. Results for the tip colder and warmer 
than the sample are illustrated in Fig. 13.

The effect of broadening on  the thermoelectric effect is
shown in Fig. 14. We assume only intrinsic asymmetry is
present, but the effect is qualitatively the same when the
thermoelectric effect originates in energy dependent 
transmission or variations in the densities of states. The 
thermoelectric current at low temperatures increases
as the broadening parameter increases. The thermoelectric voltage
is reduced in the presence of broadening: the effect is most
dramatic when the tip is colder than the sample, the large
thermoelectric voltages obtained in the absence of broadening at low
temperatures are significantly reduced and become comparable to
those obtained for opposite sign of the temperature gradient.
This occurs because in the presence of broadening the thermoelectric current
at low temperatures
is dominated by low-energy electrons rather than electrons with energies
above the energy gap.

Next we consider the situation where both intrinsic asymmetry
and energy-dependent transmission
 exist. As the tip-sample distance
is varied one or the other effect could dominate, as previously discussed.
Results are shown in Fig. 15 and 16.
For sufficiently small tip-sample distance, assuming tunneling remains ideal,
the intrinsic asymmetry effect will dominate at low temperatures. Here we have
assumed a gap slope $\Delta'=0.24$ and an average work function
$\bar{\phi}=1eV$. According to Eqs. (35) and (30b), the cross-over between
intrinsic-asymmetry dominated and barrier dominated thermoelectric effect
at low temperatures should occur for tip-sample distance $d\sim 15A$. Indeed, intrinsic
asymmetry is seen to dominate for $d=5A$ and $d=10A$. Note that the
cross-over in the thermoelectric voltage at low temperatures  
when the tip is colder than the sample (Fig. 16(a)) would be very sharp
as function of tip-sample distance. Note also that in the tunneling conductance
instead the cross-over between gap-slope dominated asymmetry and barrier dominated
asymemtry would instead occur only for $d\sim 30A$ for these parameters,
i.e. the intrinsic asymmetry would dominate for a much larger range
of tip-sample distances.
Observation
of the behavior discussed here  would clearly evidence the competition
between intrinsic and barrier-induced effects and allow for the
extraction of the intrinsic gap slope.

Finally we compare in Figure 17 the behavior of tunneling
asymmetry and thermoelectric voltage as function of tip-sample distance,
in the presence of intrinsic asymmetry, barrier-induced asymmetry,
and energy-dependent sample density of states. As discussed earlier,
intrinsic asymmetry dominates in the tunneling asymmetry over a 
substantially larger range of tip-sample distances than in the
thermoelectric effect. For both cases, broadening increases the
range of tip-sample distances where intrinsic asymmetry dominates.

\section{Discussion}

We have discussed in this paper various factors that may give rise to
asymmetry in STM tunneling conductance experiments. While the various
effects discussed here can give rise to asymmetries of either sign, we
have argued that the recent results on tunneling spectra of $Bi_2Sr_2CaCu_2O_{8+\delta}$
provide  evidence for an asymmetry originating in an intrinsic
property of the superconductor, an energy-dependent gap function. 
The slope of the gap function as function of electronic energy implied by
the experimental results is $positive$. We have suggested
that further data on temperature and tip-sample dependence of the 
asymmetry may be able to further support this conclusion.

Furthermore we have proposed that independent evidence for intrinsic
asymmetry originating in a finite gap slope
would be provided by thermoelectric experiments with STM,
which should yield $positive$ thermopower in certain parameter
regimes.
Such experiments, always yielding $negative$ thermopower,
 have already been performed with normal 
metals, which suggests that they are quite feasible.

The combination of more extensive high quality tunneling spectra and
results for the thermoelectric effect should be able to determine unambiguously
the sign and magnitude  of the gap slope in $Bi_2Sr_2CaCu_2O_{8+\delta}$.
The gap slope determines the average quasiparticle  charge in the
superconductor and is thus a quantity of fundamental interest. The theory of hole 
superconductivity\cite{hole1} has predicted that the gap slope, and 
consequently also the average quasiparticle charge, is positive for all
superconductors, and that its magnitude scales with the critical 
temperature. Temperature and carrier concentration dependence of the gap slope
is also predicted by the theory.

Coffey and coworkers\cite{coff} have recently proposed that the observed
asymmetry in tunneling conductance discussed here is evidence in favor of
d-wave symmetry of the order parameter. Their analysis relies on an
assumption of directional tunneling, and the parameters in their model are
chosen in order to match the observed asymmetry. We note that for the
case of a half-filled band and a band structure with only nearest neighbor
coupling, that is an electron-hole symmetric system, their analysis predicts 
an asymmetry of the observed sign when the tunneling direction is parallel
or close to one of the principal axis in the plane for both a d-wave and an
s-wave gap. Even though the asymmetry predicted is larger for d-wave, the
essential element giving rise to the asymmetry in this calculation appears
to be the assumption of directional tunneling. While this assumption may 
have validity for the case of point contact tunneling it would appear not to
be applicable to the STM experiments discussed here where the tip is
mounted perpendicular to the Bi-O layers and tunneling is expected to yield
an angular average over the ab-plane density of states\cite{stm1}.

Recent STM experiments in the presence of a magnetic field\cite{renn2}
provide further strong evidence for the existence of an intrinsic
asymmetry associated with the superconducting state. The data of 
Renner et al clearly show that when the STM tip is moved from
the vortex core to a region between vortices the peak for negative
sample voltages grows faster than that for positive voltages, 
irrespective of what the sign of the asymmetry in the vortex core was.
This strongly suggests that the asymmetry is directly associated
with the superconducting state. As we have discussed, the opening of a
gap in the absence of intrinsic asymmetry would lead to precisely
the opposite effect, i.e. the positive voltage peak growing faster,
due to the energy dependence of transmission. Further support for this point is
provided by the fact that Renner et al state that 'the sharpest contrast
to map the vortex structure on BSCCO is obtained using the conductivity at
a negative sample voltage $V=-\Delta_p/e$'; ($\Delta_p=$ peak position. In
other words, the peak at negative voltages gives the strongest signal on
whether the system is normal or superconducting. This is
related to the prediction of the theory of hole superconductivity that
photoemission should give a much clearer signal of the transition to
the superconducting state than inverse photoemission\cite{photo}.
Similarly one should
find a change in the sign of the thermoelectric voltage as function of
the tip position in measurements in the mixed state, with negative 
thermopower measured in the vortex core region and positive thermopower in regions
far from the cores.

Future work will involve a calculation of tunneling spectra as function
of position in the vortex lattice within a model with finite gap
slope such as the model of hole superconductivity. 
Possible signatures of a finite gap slope 
in Josephson tunneling are being investigated. It would
be of great interest to find other experimental signatures of a finite
gap slope in superconductors.

\acknowledgements
The author is grateful to Z. Yusof and J.F. Zasadzinski for clarifying 
correspondence on Ref. 13.

\begin{figure}
\caption { 
Diagram of electron energy levels. $V$ is the voltage of the sample
relative to the tip, $E$ is the electronic energy above the
Fermi level of the tip. $\phi_t$ and $\phi_s$ are the work
functions of tip and sample.
}
\label{Fig. 1}
\end{figure}

\begin{figure}
\caption { 
Tunneling conductance for a superconductor with zero temperature
gap $\Delta=30meV$ and $T_c=100K$, for $T/T_c=0.1, 0.25, 0.5, 0.85$ and $1$.
Constant densities of states of sample and tip and energy-independent
transmission is assumed. In the following figures, the same values for
gap and temperatures are used. (a) $\Gamma=0$, (b) $\Gamma=3meV$.
}
\label{Fig. 2}
\end{figure}

\begin{figure}
\caption { 
Effect of finite barrier height. Barrier thickness is $d=20A$. In (a),
work functions of tip and sample are equal. The conductance is symmetric
in the voltage above $T_c$, and becomes asymmetric as the gap opens.
In (b), the conductance above $T_c$ is asymmetric due to unequal work
functions; as the gap develops, the asymmetry switches sign.
Broadening parameter is $\Gamma=3meV$.}
\label{Fig. 3}
\end{figure}

\begin{figure}
\caption { 
Illustration of asymmetry in tunneling resulting from energy gap in the sample.
For equal work functions, 
the transmission will be higher for positively biased sample because
the tunneling electrons are closer to the top of the barrier.
}
\label{Fig. 4}
\end{figure}

\begin{figure}
\caption { 
Effect of non-constant density of states. The energy variation of the
normal state density of states is linear, and given by the conductance
curve for $T=T_c$. It corresponds to either a sample density of states 
that decreases with electronic energy, or a tip density of states that
increases with energy, or a combination of the two. The asymmetry
decreases as the temperature is lowered, both without (a) and
with (b) broadening. The thin lines connecting the peaks are
drawn to illustrate this effect.}
\label{Fig. 5}
\end{figure}

\begin{figure}
\caption { 
Effect of sloped density of states (a) and of bandwidth cutoff (b)
together with energy-dependent transmission. $\phi_s=\phi_t=1eV$,
barrier thickness $d=10A$. $\Gamma=3meV$. The dashed line in (b)
shows the density of states of the sample.}
\label{Fig. 6}
\end{figure}

\begin{figure}
\caption { Effect of intrinsic asymmetry, without (a) and with (b) broadening.
Gap slope is $\Delta'=0.12$. Constant densities of states and
energy-independent transmission is assumed.
}
\label{Fig. 7}
\end{figure}

\begin{figure}
\caption { Effect of intrinsic asymmetry and energy-dependent transmission.
Gap slope is $\Delta'=0.12$, barrier parameters are $\phi_s=\phi_t=1eV$,
$d=10A$. (a) Constant density of states, (b) sloped density of states.
}
\label{Fig. 8}
\end{figure}

\begin{figure}
\caption { Tunneling characteristics in the presence of non-constant density
of states, intrinsic gap slope and energy-dependent transmission, at temperature
$T/T_c=0.1$ for various tip-sample distances given in the figure (numbers next
to the curves). 
The curves are offset increasingly by $0.5$ for clarity. Broadening
parameter is $\Gamma=3meV$. Gap slope is $\Delta'=0.17$ and work functions
are $\phi_s=\phi_t=0.53 eV$.
}
\label{Fig. 9}
\end{figure}

\begin{figure}
\caption { Current versus voltage for sample and tip at different
temperatures as given in the figure for various cases,
 with constant densities of states, zero
gap slope and energy-independent transmission except as indicated below.
Curve labeled a (solid line): gap slope $\Delta'=.12$; b (dashed line): 
tip density of states of slope $m_t/g_t(0)=-4eV^{-1}$; c (dot-dashed line):
sample density of states of slope $m_s/g_s(0)=4eV^{-1}$; d (solid line):
barrier parameters $\phi_s=\phi_t=0.5eV, d=20A$, and gap slope $\Delta'=0.36$.
No broadening is assumed.
}
\label{Fig. 10}
\end{figure}

\begin{figure}
\caption { Tunneling characteristics for the parameters of the four cases
in Figure 10. (a) sloped gap , (b) sloped tip
density of states, (c) sloped sample density of states, (d) sloped gap and
energy-dependent transmission. Note that the sloped densities of states
of sample and tip give rise to tunneling conductance of opposite slope but
to thermoelectric effect of the same sign (Fig. 10). Note also that
the tunneling asymmetry in (d) is of the sign expected from the gap
slope, but the sign of the thermoelectric effect in Fig. 10 (curve labeled d)
is opposite 
because the effect of energy-dependent transmission dominates there. 
}
\label{Fig. 11}
\end{figure}

\begin{figure}
\caption { Zero voltage thermoelectric current from sample to tip
for the tip colder than the sample as function of sample temperature
for fixed temperature difference between tip and sample.
Constant densities of states, zero
gap slope and energy-independent transmission is assumed except as indicated below.
Solid line: gap slope $\Delta'=.12$; Dashed line: 
tip density of state of slope $m_t/g_t(0)=-4eV^{-1}$; Dotted line:
sample density of state of slope $m_s/g_s(0)=4eV^{-1}$; Dot-dashed line:
barrier parameters $\phi_s=\phi_t=1eV, d=7.84A$. $\Gamma=0$ in all cases.
}
\label{Fig. 12}
\end{figure}

\begin{figure}
\caption { Zero current thermoelectric voltage (of sample) for the four cases
of Fig. 12. The line convention is the same as in Fig. 12. (a) Tip colder than
the sample. Note that the voltage becomes very large at low temperatures. The inset shows
the behavior of the voltage close to $T_c$. (b) Sample colder than tip. Note that
the voltages are much smaller than in (a). Note also that both in (a) and (b)
only in the case of
intrinsic asymmetry alone (solid line) does the voltage go to zero at $T_c$.
}
\label{Fig. 13}
\end{figure}

\begin{figure}
\caption { 
Effect of broadening on thermoelectric current (a) and 
thermoelectric voltage (b). Gap slope is $\Delta '=0.12$, constant
densities of states and energy-independent transmission is assumed. 
Solid lines are for $\Gamma =0$, dashed lines for $\Gamma =1meV$ and
dash-dotted lines for $\Gamma =3meV$. The effect of
broadening is to  slightly  increase the thermoelectric current.
The thermoelectric voltage is somewhat decreased by broadening when
the tip is hotter than the sample, and substantially so
when the tip is colder than the sample.
}
\label{Fig. 14}
\end{figure}

\begin{figure}
\caption {  Zero voltage thermoelectric current (from sample to tip)
versus sample temperature in the 
presence of both finite gap slope, $\Delta'=0.24$, and energy-dependent transmission
with barrier parameters $\phi_s=\phi_t=1eV$ and tip-sample distances given
next to the curves. Since the sample is colder than the tip, positive thermopower
corresponds to current going from tip to sample, which occurs when the barrier is 
thin and the
intrinsic asymmetry dominates. For temperature gradient of opposite sign the
curves would be the mirror images of these across the horizontal axis. $\Gamma=0$.
}
\label{Fig. 15}
\end{figure}

\begin{figure}
\caption { Zero current thermoelectric voltage (of sample) for parameters as
in Figure 15 and (a) tip colder than sample and (b) tip hotter than sample. 
Thermoelectric power is positive for thin barriers and low temperatures.
The inset in (a) shows the behavior close to $T_c$.
}
\label{Fig. 16}
\end{figure}

\begin{figure}
\caption { Thermoelectric voltage (of tip)  with $\Gamma=0$
(solid lines) and with $\Gamma=3meV$ (dashed lines) at two temperatures
 (labels next to the lines) for $T_t-T_s=-0.1 T_c$ as function of
tip-sample distance $d$. Sample density of states is the same as in Fig. 5.
 The data for $\Gamma=3meV$ are multiplied
by $50$.
For small $d$ intrinsic asymmetry dominates, for large $d$ barrier-induced
asymmetry dominates. 
Note that broadening causes the thermoelectric effect to be
dominated by intrinsic asymmetry over a wider range of tip-sample
distances. 
The tunneling asymmetry at temperature $T=0.1T_c$ is also shown
for $\Gamma=0$ and  $\Gamma=3meV$ (dash-dotted lines labeled $A$).
Here again, the asymmetry remains dominated by the intrinsic effect
for a larger $d$ in the case where $\Gamma$ is finite.
Gap slope is $\Delta'=0.17$, work functions are $\phi_s=\phi_t=0.53eV$.
 Note that the range of barrier thickness where intrinsic
asymmetry dominates is substantially larger for the tunneling conductance asymmetry
than for the thermoelectric effect. 
}
\label{Fig. 17}
\end{figure}


\begin{references}
\bibitem{nega}Note that the conductance becomes negative in Fig. 6(b) for
positive voltages. The fact that negative conductance can occur in
STM in the presence of rapidly varying density of states is well
understood and has in fact been observed experimentally\cite{book}.
\end{references}
\end{document}